\documentstyle[aps,pre,epsfig,preprint]{revtex}

\title{Electrophoresis simulated with the cage model for reptation}

\draft

\author{H.R.W. Beljaars \and A. van Heukelum\cite{correspondent}}

\address{Theoretical Physics, Utrecht University,\\
Princetonplein 5, 3584 CC Utrecht, the Netherlands}

\date{\today}

\begin{document}

\maketitle

\begin{abstract}
The cage model for polymer reptation is extended to simulate DC
electrophoresis. The drift velocity $v$ of a polymer with length
$L$ in an electric field with strength $E$ shows three different
regions: if the strength of field is small, the drift velocity
scales as $v\sim E/L$; for slightly larger strengths, it scales
as $v\sim E^2$, independent of length; for high fields, but
still $E\ll 1$, the drift velocity decreases exponentially to
zero. The behaviour of the first two regions are in agreement
with earlier reports on simulations of the Duke-Rubinstein model
and with experimental work on DNA polymers in agarose gel.

Report no.: THU-99/13
\end{abstract}

\pacs{82.45.+z, 05.40.+j, 87.15.He}

\section{Introduction}

A widely used tool to separate mixtures of DNA molecules by
length is DC electrophoresis. The DNA is confined to an agarose
gel, and an electric field is applied. Since DNA is negatively
charged, it moves towards the positive electrode as a result of
this electric field. As the drift velocity depends on the
length, DNA fragments with different lengths end up in different
bands, and can therefore easily be separated.

\begin{figure}
\caption{Movement of a defect along the chain. When a defect
moves along the chain, it displaces monomers which it passes by
a distance equal to the stored length.}
\label{figdegennes}
\noindent
\epsfig{file=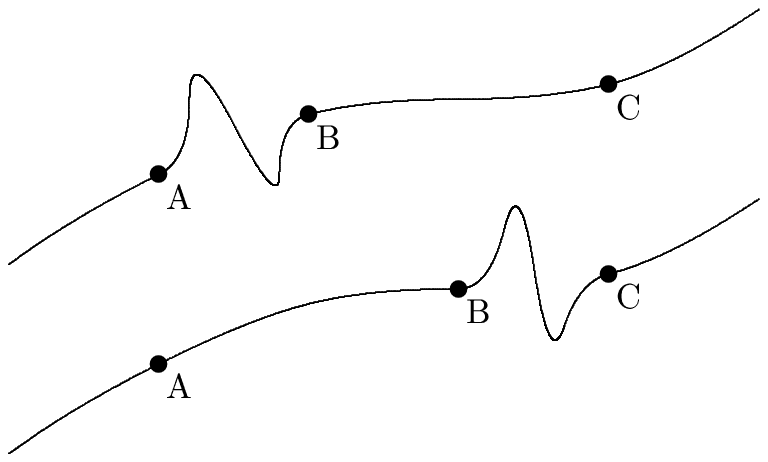, width=3.0in}
\end{figure}

Since DNA fragments are usually much longer than the typical
spacing between the gel strands, they are unable to move
sideways. De Gennes \cite{Degennes71} described the motion of a
polymer in such an environment, and termed it {\em reptation}:
the polymers move by diffusion of `defects' along the chain of
monomers. Each defect contracts the polymer by a certain amount
of length, called its {\em stored length}. When a defect passes
a monomer, the monomer is moved by this distance. Figure
\ref{figdegennes} shows an example where a defect travels past
monomer $\mbox{B}$.

Two models are widely used to simulate reptation: the repton
model, introduced by Rubinstein \cite{Rubinstein87}, and the
cage model, introduced by Evans and Edwards \cite{Evans81}. In
both models, monomers reside on sites of a simple cubic lattice
(or in two dimensions a square lattice), and are connected by
bonds; the dynamics consist of single-monomer moves.

In the repton model, stored length consists of zero-length
bonds. For this model, it was already proposed by Rubinstein
\cite{Rubinstein87}, and later proven by Pr\"ahofer and Spohn
\cite{Prahofer96} that the diffusion constant $D$ of the
polymer in the limit of long polymer length $L$ obeys the
scaling $L^2 D= 1/3$. For finite lengths, the diffusion constant
is known numerically exact up to length 20 and from Monte Carlo
simulations up to length 250 \cite{Newman97}. The repton model
has been adapted for the study of DC electrophoresis by Duke
\cite{Duke89}. This Rubinstein-Duke model has been studied
numerically for lengths up to 400 \cite{Barkema94}. Simulations
of this model are easy because it can, without loss of
generality, be reduced to a one-dimensional model.

In the cage model, stored length consists of a pair of
anti-parallel nearest-neighbour bonds, called `kinks'. The
polymer diffusion constant in this model has been determined
numerically exact for small $L$ \cite{Heukelum99}, and with
Monte Carlo simulations for polymers up to length 200
\cite{Deutsch1989x91,Barkema98}. As in the repton model, the
polymer diffusion constant scales as $D\sim L^{-2}$. In this 
paper, we extend the cage model to simulate a charged polymer
in an electric field. Interestingly, we find differences in the 
scaling of the polymer drift velocity as compared to the 
Duke-Rubinstein model.

In section \ref{CageModel} we describe the cage model and
present how the model can be extended to simulate reptation in a
non-zero electric field. We discuss in sections \ref{multispin}
to \ref{multispinlast} how efficient simulations can be achieved
with multispin coding; these sections are not needed for
understanding other parts of the paper. In section
\ref{Simulations} we discuss the simulation approach in detail.
In section \ref{secVelocity} we discuss scaling arguments for 
the drift velocity. The results are presented in section 
\ref{Results} which includes statements about the polymer
shapes and comparisons to previous reports.

\section{Cage model}\label{CageModel}

The cage model describes a polymer of $L$ monomers, located on
the sites of an infinite cubic lattice. The monomers are
connected by $L-1$ bonds with a length of one lattice spacing. A
single step of the Monte Carlo simulation consists of selecting
randomly a monomer and, if it is free to move, moving it to a
randomly selected location (possibly the current location).

The monomers at both ends of the polymer are always free to
move, but monomers in the interior of the polymer are only free
to move when the two neighbours along the chain are located on
the same adjacent lattice site. Other movements might result in
an acceptable polymer configuration, but are ruled out because
they would allow the polymer to move sideways, which is not
reptation. One possible interior move is shown in figure
\ref{Polymer}. Every possible move occurs statistically with
unit rate, setting the time scale. A single elementary move thus
corresponds to a time increment of $\Delta t= 1/(2dL)$, where
$d$ is the dimensionality of the lattice; in our case, $d= 3$.

\subsection{Electric field}

In solution, DNA becomes negatively charged with a fixed charge
per unit length. We incorporate this into the cage model by
assigning a negative charge $q$ per monomer. The polymer is
located in a homogeneous electric field $\vec E$, that acts on
these charged monomers.

For two monomer positions $\vec{r}_1$ and $\vec{r}_2$, separated
by a displacement $\vec{r}_{12}= \vec{r}_2-\vec{r}_1$, the
difference in potential energy is given by
$U= q\vec{E}\cdot\vec{r}_{12}$. The ratio of the corresponding
Boltzmann probabilities is
\begin{equation}
P_1/P_2= e^{U/(k_{B}T)}= e^{{qEr}/(k_{B}T)}\mbox{,}
\end{equation}
where $E= |\vec{E}|$ is the electric field strength, and
$r= {\vec{E}\cdot\vec{r}_{12}}/{|\vec{E}|}$ is the displacement
parallel to the field.

In a Monte Carlo simulation, this ratio determines at which
rates the monomers are to be moved along the field or against
it. We choose the direction of the electric field along one of
the body diagonals of the unit cubes, because then the $x$, $y$
and $z$ directions are equivalent, and within one elementary
move, the displacement $r$ takes only the two values
$\pm{}2/\sqrt{3}$ times the lattice spacing. For convenience,
the units are chosen in such a way that $qr/k_{B}T= \pm{}1$.

Each monomer moves with a rate $R^{+}= \exp(E)$ for moves which
lower the energy and $R^{-}= \exp(-E)$ for moves which raise the
energy. When a monomer is selected and is able to move to lower
(higher) energy, it will do so with a probability $P^{+}$
($P^{-}$), given by
\begin{equation}
P^{+}= \frac{1}{d}~\frac{e^{E}}{e^{E}+e^{-E}}\mbox{,}\quad
P^{-}= \frac{1}{d}~\frac{e^{-E}}{e^{E}+e^{-E}}\mbox{.}
\label{ProbabilityDirection}
\end{equation}

The time increment corresponding to one elementary Monte Carlo
move is thus equal to
\begin{equation}
\Delta{}t= \frac{1}{dL}\frac{1}{e^{E}+e^{-E}}\mbox{.}
\end{equation}

\subsection{Bond Representation}\label{BondRep}

\begin{figure}
\caption{One elementary move of a monomer: a `kink' (pair of
anti-parallel neighbouring bonds) is replaced by another one.}
\label{Polymer}
\noindent
\epsfig{file=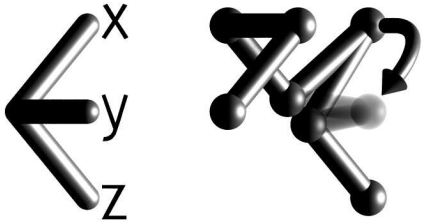, width=2.4in}
\end{figure}

As described in section \ref{CageModel} the monomers are
connected by bonds, where each bond has one of $2d$ possible
orientations. One way of describing the polymer configuration is
by specifying the location of the first monomer and the
orientation of all bonds. The advantage of this notation is that
only the position of one monomer has to be stored plus the
orientations of all bonds. The polymer in figure \ref{Polymer},
for example, is described by the position of the first monomer,
on the left side of the figure, and \verb|+x-y+z+x-x+z|.

The dynamics can be described in terms of bonds. The bonds that
are located on both ends of the polymer are always free to move.
The internal bonds are free to move only when they are part of a
pair of oppositely oriented neighbouring bonds (a kink). The
first and last monomer in figure \ref{Polymer} can change to any
new bond: \verb|+x|, \verb|+y|, \verb|+z|, \verb|-x|, \verb|-y|
or \verb|-z|. The kink configuration \verb|+x-x| can change in
any new kink: \verb|+x-x|, \verb|+y-y|, \verb|+z-z|,
\verb|-x+x|, \verb|-y+y| or \verb|-z+z|.

\subsection{Multispin Coding}\label{multispin}

With multispin coding, many polymers can be simulated in
parallel. We used an approach similar to the one by Barkema and
Krenzlin \cite{Barkema98}. The simulation that we performed used
64-bit unsigned integers to simulate 64 different polymers in
parallel. As described in section \ref{BondRep}, there are six
directions a bond can point to, so each bond can be encoded
using three binary digits. It is now possible to encode 64 bonds
in three integers $x$, $y$ and $z$, as shown in table
\ref{bondencoding}.

In each iteration of the inner loop of the algorithm a random
monomer $i$, $0 \leq i < L$, is selected. When an inner monomer
is selected the two surrounding bonds are compared; if they are
opposites, they are replaced by a randomly generated pair of
opposite bonds. Section \ref{multispinlast} describes how to
generate those bonds. The first and last monomers are special
cases, which are described in section \ref{firstlast}.

To find the kinks in all of the 64 polymers, we use Equation
(\ref{findkink}):
\begin{equation}\label{findkink}
\begin{array}{rcl}
k_i&=&(x_{i-1} \oplus x_i)\\
&\wedge&(y_{i-1} \oplus y_i)\\
&\wedge&(z_{i-1} \oplus z_i)\mbox{.}
\end{array}
\end{equation}
Monomer $i$ is surrounded by bonds $i-1$ and $i$. Bit $j$ of
$k_i$ is $1$ if the surrounding bonds of monomer $i$ of polymer
$j$ are in opposite directions.

If a monomer can be moved, it will be relocated using a list of
random kinks encoded in $\hat x$, $\hat y$ and $\hat z$. Bonds
$i-1$ and $i$ that surround monomer $i$ are replaced by
$\hat x$, $\hat y$ and $\hat z$ and their binary complements,
respectively. Equation (\ref{replacekink}) shows how this can
be done:
\begin{equation}\label{replacekink}
\begin{array}{rcl}
x_{i-1}&=&(\neg k_i \wedge x_{i-1}) \vee (k_i \wedge\hat x)\\
y_{i-1}&=&(\neg k_i \wedge y_{i-1}) \vee (k_i \wedge\hat y)\\
z_{i-1}&=&(\neg k_i \wedge z_{i-1}) \vee (k_i \wedge\hat z)\\
x_i&=&(\neg k_i \wedge x_i) \vee (k_i \wedge \neg\hat x)\\
y_i&=&(\neg k_i \wedge y_i) \vee (k_i \wedge \neg\hat y)\\
z_i&=&(\neg k_i \wedge z_i) \vee (k_i \wedge \neg\hat z)\mbox{.}
\end{array}
\end{equation}

These 27 logical operations replace the kinks near monomer $i$
in all 64 polymers; polymers that have no kink near monomer $i$
are left unaltered.

\subsection{First and last monomer}\label{firstlast}

The first and last monomers are always free to move. When one of
those monomers is selected we can just replace the bonds with
randomly generated bonds. To keep track of the position of the
first monomer we have to calculate the distances traveled in the
$x$, $y$ and $z$ directions. Since those directions are
symmetric we only calculate $r = x+y+z$. For this we only need
to know whether the first bonds point at a negative direction,
which is one of \verb|-x|, \verb|-y| and \verb|-z|. This is done
using the following equation:
\begin{equation}\label{negativebonds}
d = (x_0 \wedge y_0)
\vee (y_0 \wedge z_0)
\vee (z_0 \wedge x_0)\mbox{.}
\end{equation}

Now the new random bonds are inserted:
\begin{equation}
\begin{array}{c}
x_0 = \neg \hat x\\
y_0 = \neg \hat y\\
z_0 = \neg \hat z\mbox{.}
\end{array}
\end{equation}
Monomer $0$ only has one bond, which is number $0$. Section
\ref{multispin} tells us that we have to use the binary
complement of the random kink.

We now calculate once more whether the first bond points at a
negative or positive direction, and with this information we can
calculate the new positions of the first monomers:
\begin{equation}
r_i = r_i - 2 d_{\mbox{\scriptsize before}}^{(i)} +
2 d_{\mbox{\scriptsize after}}^{(i)}\mbox{.}
\end{equation}

When the last monomer is selected, it can simply be replaced
with random new bonds:
\begin{equation}
\begin{array}{c}
x_{L-2} = \hat x\\
y_{L-2} = \hat y\\
z_{L-2} = \hat z\mbox{.}
\end{array}
\end{equation}

\subsection{Generation of random kinks}\label{multispinlast}

The algorithm described above relies on the availability of
random kinks. These kinks should be generated with the
probabilities as given in equation (\ref{ProbabilityDirection}).
Since the two bonds in a kink have opposite directions only one
bond has to be generated; the bond on the other side of the
monomer is easily derived.

The properties of detailed balance are used to create those
bonds correctly. Certain properties must be enforced: first of
all the \verb|x|, \verb|y| and \verb|z| directions should occur
with the same probability; secondly the ratio of the
probabilities for \verb|+| and \verb|-| bonds is given by
quotient of $P^{-}$ and $P^{+}$, as given in equation
(\ref{ProbabilityDirection}); this quotient is given by
\begin{equation}
P^{\mbox{\scriptsize rel}} = P^{+}/P^{-} = e^{-2 E}\mbox{.}
\end{equation}

The first property is enforced by rotating some of the bonds (we
used 50\%) the following way: \verb|x|$\mapsto$\verb|y|,
\verb|y|$\mapsto$\verb|z| and \verb|z|$\mapsto$\verb|x|. Using
the randomly generated bit pattern $r$ the following statements
are used to rotate the bonds:
\begin{equation}
\begin{array}{l}
\tilde x=(r \wedge\hat x) \vee (\neg r \wedge\hat y)\\
\tilde y=(r \wedge\hat y) \vee (\neg r \wedge\hat z)\\
\tilde z=(r \wedge\hat z) \vee (\neg r \wedge\hat x)\mbox{.}
\end{array}
\end{equation}

The second property is then enforced by inverting some of the
bonds. With 50\% probability, the negative bonds are inverted,
with $P^{\mbox{\scriptsize rel}}$ times 50\%, also the positive
bonds are inverted. To make sure that all random kinks are
independent we create a list of those and reshuffle this list
regularly.

\section{Simulations}\label{Simulations}

The simulation algorithm described in sections \ref{multispin}
to \ref{multispinlast} was implemented using the C programming
language. The random number generator we used is a lagged
(24, 55) additive Fibonacci generator. The simulations are done
on a Silicon Graphics Origin 200 (180 Mhz) and on a DEC Alpha
(466 Mhz) computer. The latter is faster and takes about 1.1
${\mu}$s per Monte Carlo step.

The polymers where initialized in a U-shape with both ends in
the direction of the electric field. At regular intervals we
checked whether a polymer has moved at least its own size, which
is the maximum distance between any two monomers. When this has
occurred for a polymer, we assume that the polymer has
thermalized, the real measurement starts when this
thermalization has finished.

The measurement is stopped when all polymers have thermalized
and the average distance traveled by all polymers is a few times
their own size. We assume that measurements are statistically
independent when a polymer has traveled a distance equal to its
own size.

We have performed simulations for lengths $3$ up to $200$. The
time taken to calculate the drift velocity varied from a few
seconds for small polymers up to about 17 hours for the longest
polymers ($L=200$) in the smallest electric field ($E=0.001$).
Simulations of longer polymers take too much time to compute the
drift velocity for small electric fields.

\section{Scaling arguments for the drift velocity}
\label{secVelocity}

The velocity of a polymer in a small electric field behaves
according to the Nernst-Einstein relation, $v=FD$, where $F=qLE$
is the force. The diffusion constant can thus be calculated from
the drift velocity by $D=v/(qLE)$ in the limit $E\to 0$. De
Gennes \cite{Degennes71} found the diffusion constant to be
proportional to $D\sim L^{-2}$. This means the drift velocity is:
$v\sim qE/L$.

For slightly larger electric fields the Nernst-Einstein relation
breaks down. Barkema, Marko and Widom \cite{Barkema94} give an
intuitive explanation of the dependence of the drift velocity on
the electric field strength. The argument goes as follows.

A random polymer will have an end-to-end length around
$h=\sqrt{L}$. When an electric field is applied, the polymer is
stretched in the direction of the electric field. When the
electric field exceeds a certain level, the polymer as a whole
does no longer resemble a random walk: $h>\sqrt{L}$. One may cut
the polymer into $n_b$ pieces (`blobs') of length $L_b= L/n_b$,
that each still look like a random walk; the average end-to-end
distance of the blobs is equal to
$\langle h_b \rangle= \sqrt{L_b}$.

Two forces work on the blob. The elastic force tries to contract
a stretched polymer and is proportional to the size of the blob
and inversely proportional to the length of the part of the
polymer that forms the blob:
$F_{\mbox{\scriptsize elastic}}\sim h_b/L_b$. The electric force
tries to stretch the polymer and is proportional to the size of
the blob as well as the electric field:
$F_{\mbox{\scriptsize electric}}\sim h_b E$. These two forces
have to be in balance which implies that the blob size is
$L_b\sim E^{-1}$.

The Nernst-Einstein relation now applies to the blobs, so
$v= F_b D_b= q L_b E D_b$. Again, if the blob size is large
enough, $D_b\sim L_b^{-2}$ which makes the speed of the polymer
quadratic in the electric field: $v\sim E/L_b\sim E^2$. This
effect has already been observed in the Rubinstein-Duke model
by Barkema, Marko and Widom\cite{Barkema94}.

\section{Results}\label{Results}

\begin{figure}
\caption{Velocity of polymers of lengths $3$ up to $200$ in
electric fields between $0.001$ and $1$. The velocity is plotted
on the vertical logarithmic axis. On the horizontal logarithmic
axis the electric field $E$ is plotted.}
\label{Velocity}
\noindent
\epsfig{file=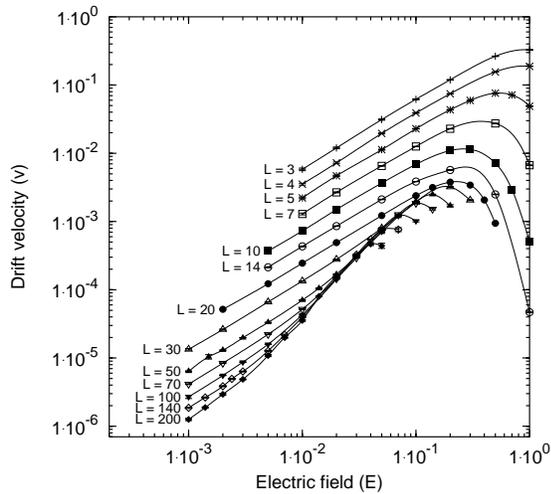, width=3.0in}
\end{figure}

The results of our simulations are presented in figure
\ref{Velocity}. The short polymers, up to length $20$, show a
behavior different from the longer polymers. These short
polymers have no superlinear dependence on the velocity on the
electric field.

\begin{figure}
\caption{Transition between the linear and quadratic dependence
of the velocity on the electric field. For various polymer
lengths, the scaled velocity $v'=(\sqrt{b}/a) v$ is plotted as a
function of scaled electric field $E'=(\sqrt{b/a}) E$, where $a$
and $b$ are $L$-dependent parameters given in table \ref{ab}.
The curve is given by $v'= \sqrt{{E'}^2+ {E'}^4}$.}
\label{resultCollapse}
\noindent
\epsfig{file=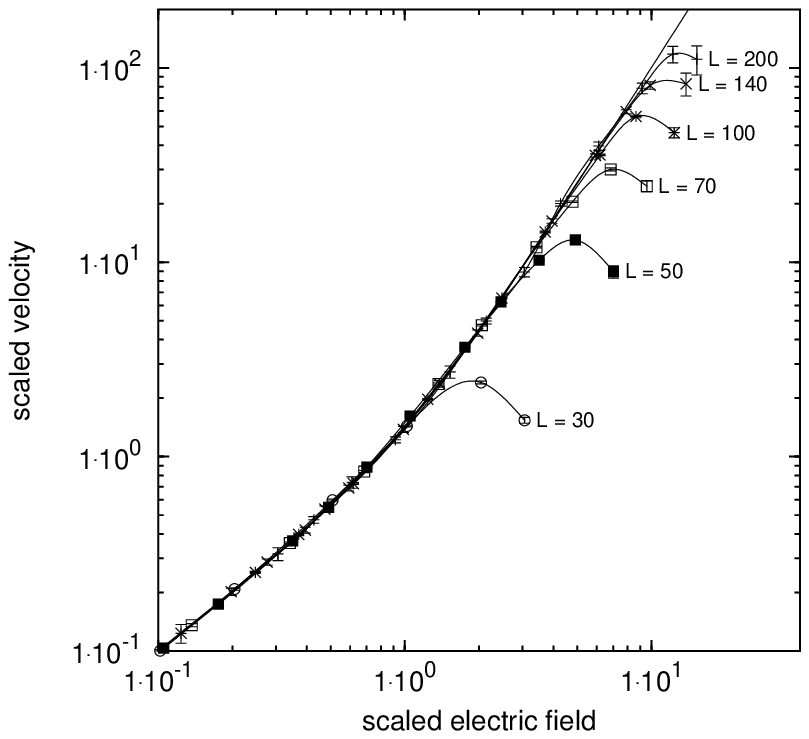, width=3.0in}
\end{figure}

When a small force, $EL\ll 1$, is applied to the polymers, the
velocity of the polymers varies linearly with the electric
field. When a force around $EL \approx 1$ is applied to the
longer polymers, the polymer velocity depends superlinearly on
the electric field. We show in figure \ref{resultCollapse} that
the dependence becomes quadratic for long polymer chains, as
derived in section \ref{Simulations}. For much larger electric
fields, the velocity decreases to zero.

\begin{figure}
\caption{The position of the center of mass of a polymer of
length $100$ divided by the electric field for the electric
fields $0.1$, $0.03$, $0.01$, and $0.003$. The lines are the
expected positions using the average speed measured by the long
experiments.}
\label{resultSim}
\noindent
\epsfig{file=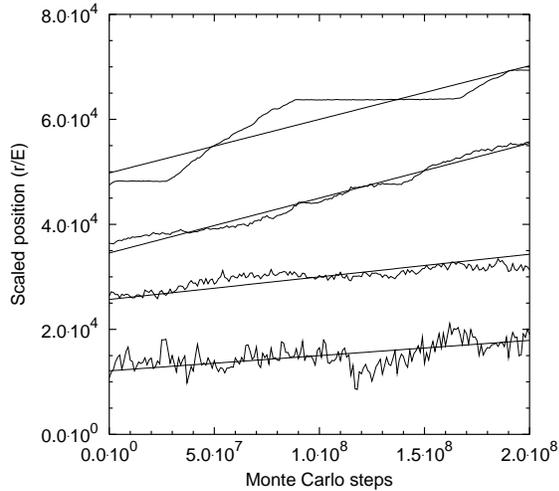, width=3.0in}
\end{figure}

For polymer length $L=100$ we performed some short simulations
to get insight in the typical movement of the center of mass of
the polymer. In figure \ref{resultSim} the position of the
center of mass, scaled with a factor of $1/E$, is plotted as a
function of time, for different field strengths. The starting
point of the polymers are chosen in a way that the graphs do not
overlap. For the smallest electric fields the movement is just
like one would expect from a diffusing particle, it moves
randomly, but with some preferential direction. For the electric
field in the middle range, the diffusion effect becomes
relatively smaller. This results in a smoother behaviour. In
high electric fields the movement of the center of mass
sometimes halts, when the force on the ends of the polymer pulls
the polymer into a U-shape. When this happens the polymer has to
untangle itself before its center of mass can move forward
again.

\subsection{Polymer shapes}

\begin{figure}
\caption{Three polymers of length $100$ in different electric
fields. From left to right: $E = 0.003$, $E = 0.01$, $E = 0.03$.
Polymers in small electric fields look like random walks. In
slightly larger electric fields the ends tend to protrude. In
high electric fields the polymer does not look like a random
walk.}
\label{lowfield}
\noindent
\epsfig{file=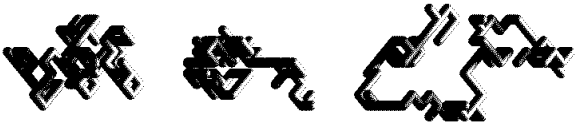, width=3.0in}
\end{figure}

The polymer shape in a small electric field resembles a random
walk, as shown on the left side of figure \ref{lowfield}. When
the electric field is increased, the shape becomes stretched
parallel to the electric field \cite{Widom1997}; the 
configuration may be viewed
as a set of blobs which move with independent speed, as
discussed in section \ref{secVelocity}. As shorter polymers move
more quickly in a given electric field, the blob-configuration
moves faster than a random walk configuration which results in a
superlinear increase of speed when the electric field is
changed.

\begin{figure}
\caption{Four polymers of length $100$, and $E = 0.1$. From top
to bottom the times in Monte Carlo steps are: $5.8\cdot 10^7$,
$8.6\cdot 10^7$, $1.25\cdot 10^8$ and $1.66\cdot 10^8$.}
\label{highfield}
\noindent
\epsfig{file=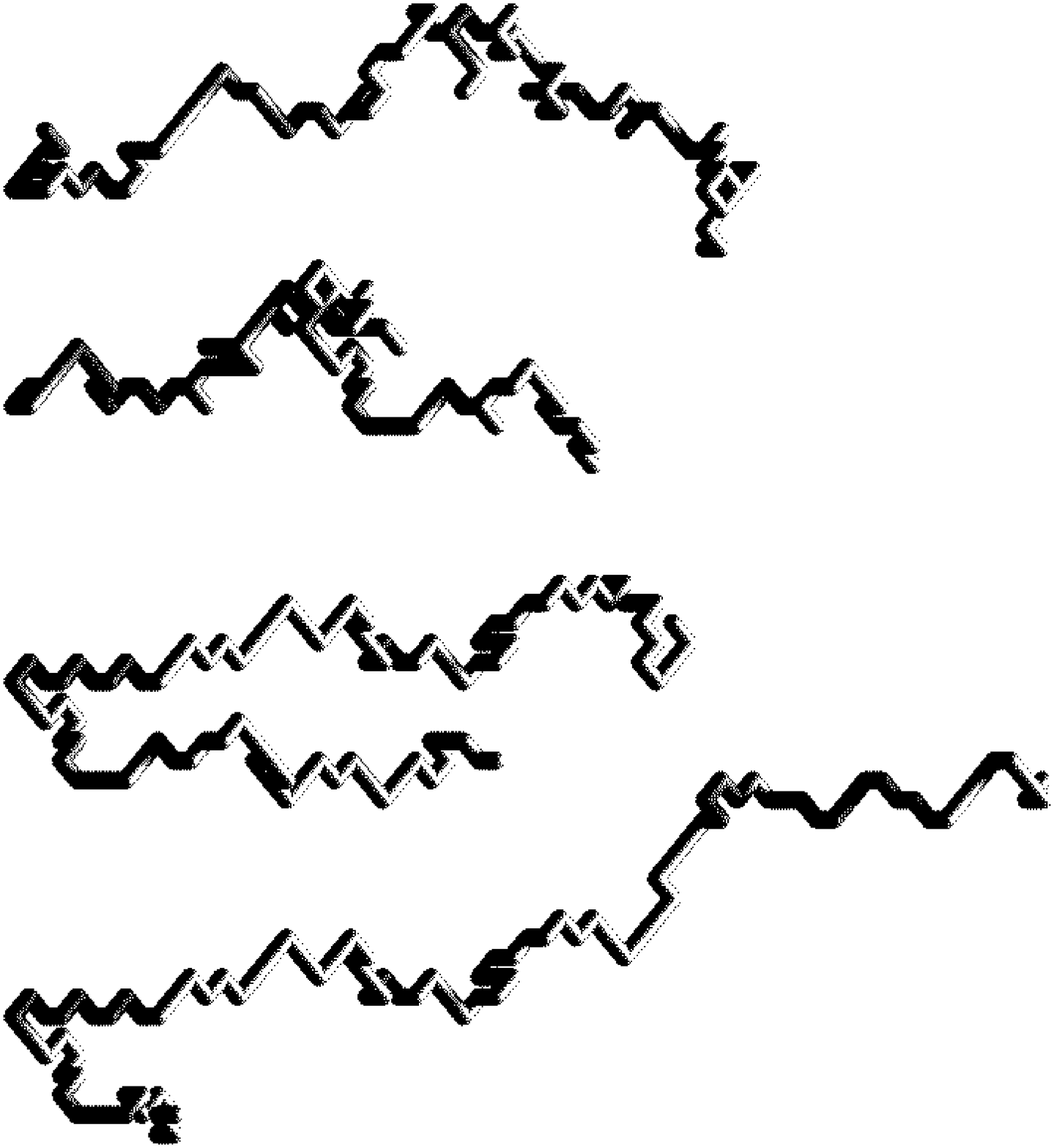, width=3.0in}
\end{figure}

When the electric field is increased above a certain value the
shape may transforms into an U-shape, as shown in figure
\ref{highfield}. With higher electric fields it becomes more
difficult to escape from this U-shape. Since the polymer cannot
move sideways it is trapped in the lattice for a long time
compared to the time it moves.

Figure \ref{highfield} shows polymers in different
configurations. The first polymer is stretched in the direction
of the electric field. This configuration may be viewed as a
large number of very small blobs. As such, the polymer has a
high velocity, which may also be seen in figure \ref{resultSim}
near $5.8\cdot 10^7$ Monte Carlo steps.

The second polymer is a transition configuration between the
fast-moving cigar-like configuration as described above and the
U-shape configuration. The polymer forms a `knot' which is later
passed by the trailing end of the polymer.

The third polymer has a typical U-shape. The polymer in this 
configuration is almost fully stretched. This means that it has
only a small number of kinks, which means that there is almost
no stored length.

Just before the polymer escapes from the U-shape, as is the
fourth polymer, its configuration is very much stretched and has
almost no stored length. This state transforms quickly in a
state that resembles the state of the first polymer in figure
\ref{highfield}.

\begin{figure}
\caption{The average number of kinks on monomer number. The
polymers are of length $100$ and electric fields are $0.003$,
$0.01$, $0.03$ and $0.1$. The line gives the expected value
$1/6$ of kinks in a random walk.}
\label{resultKinks}
\noindent
\epsfig{file=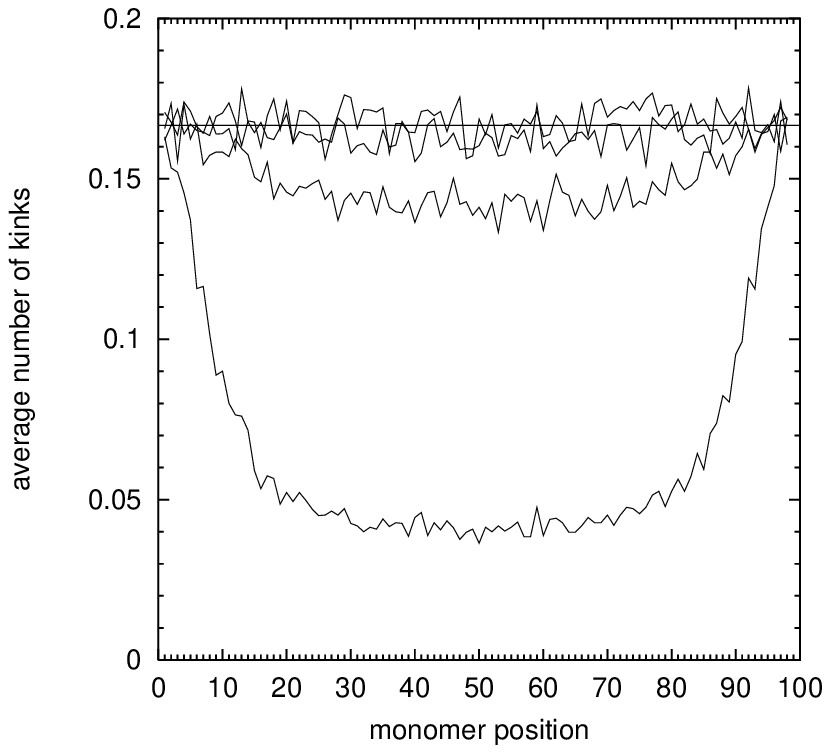, width=3.0in}
\end{figure}

As described before, the number of kinks in a polymer decreases
when the polymer is stretched. To check the dependence on the
electric field we have performed some short simulations to find
the average number of kinks on each location along the polymer.
The simulations consisted of $10^9$ Monte Carlo steps after
$2\cdot 10^8$ steps of thermalization, starting with a random
configuration. Every $10^6$ Monte Carlo steps the kinks are
counted. The fraction of time that a kink exists on a certain
location is displayed in figure \ref{resultKinks}.

For small electric fields the polymer configuration is known to
resemble a random walk in three dimensions. The average number
of kinks is thus expected to be $1/6$. For higher electric
fields the U-shape configuration becomes more frequent. In this
configuration the kinks are likely to diffuse towards the ends
of the polymer, which means that the average number of kinks in
the middle of the polymer decreases. When this happens we can no
longer apply the blob argument as described in section
\ref{secVelocity}. The mobility of the blobs in the middle of
the polymer decreases as the average number of kinks in that
region decreases.

For a longer U-shaped polymer it becomes more difficult to
escape from this configuration. The probability of a kink moving
from one end of the polymer to the other end decreases with the
length of the polymer. This means that longer polymers spend a
longer time in the U-shape configurations.

When the density of kinks becomes less than $1/6$ per monomer,
the entropic force that contracts the polymer is no longer in
balance with the electric force. The polymer itself now
transports the force along the chain, which may be better
explained by the continuous model of Deutsch and Madden
\cite{Deutsch1989x90}.

\subsection{Comparisons to previous reports}
\label{compareprevious}

The results of the Duke-Rubinstein model have been compared to
actual experiments\cite{Barkema96}. For longer polymers, the
data is well described by
\begin{equation}\label{barkema}
\frac{L^2 v}{\alpha} =
\left[ \left( \frac{L E}{\beta} \right)^2 +
\left( \frac{L E}{\beta} \right)^4 \right]^{1/2}
\end{equation}
This function is equivalent to the function $v^2= aE^2+ bE^4$,
where $a$ and $b$ are functions of $\alpha$, $\beta$ and $L$. To
check whether our results show the same scaling behaviour, we
collapsed our data to the function $v'= \sqrt{{E'}^2+{E'}^4}$ in
figure \ref{resultCollapse}, where $v'=(\sqrt{b}/a) v$ and
$E'=(\sqrt{b/a}) E$. The data in the third region is discarded
in calculating $a$ and $b$.

Experiments have been performed on DC electrophoresis
\cite{Heller94,Hervet87}, both articles confirm the existence of
regions where $v\sim E$ and $v\sim E^2$.

\begin{figure}
\caption{Diffusion constant calculated from our measurements,
compared to the scaling relation found by Barkema and Krenzlin.
This scaling relation is a straight line when $N^2 D$ is plotted 
as a function of $N^{-2/3}$.} 
\label{comparefigbarkema} 
\noindent
\epsfig{file=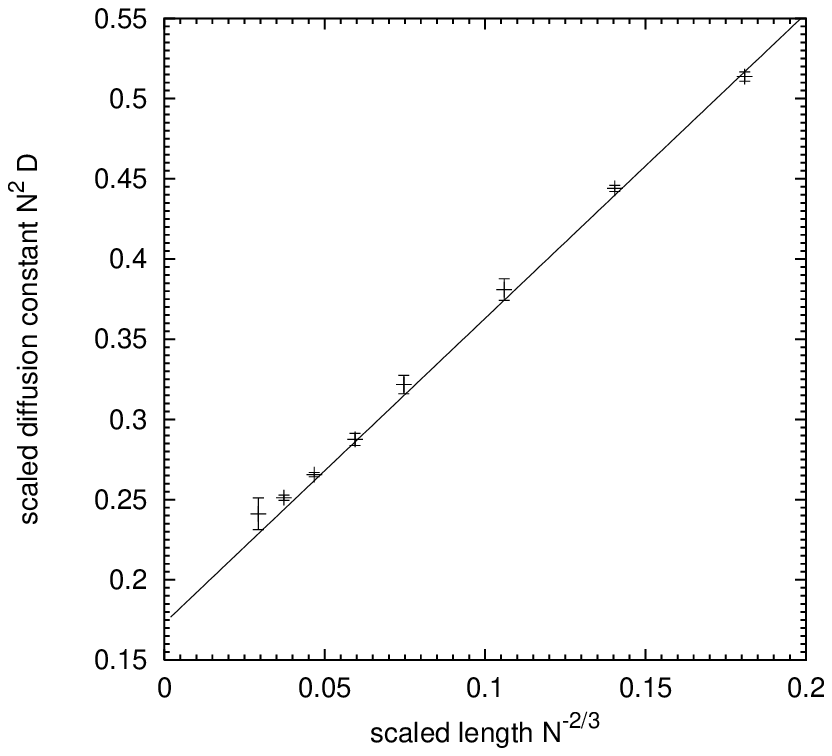, width=3.0in}
\end{figure}

The diffusion constant is equal to $D= \sqrt{a}/L$. The scaling
found by Barkema and Krenzlin \cite{Barkema98} is given by
$DN^2= 0.173+ 1.9 N^{-2/3}$, where $N= L-1$. Figure
\ref{comparefigbarkema} shows our results compared to their
scaling function. Our results agree within statistical errors.

\section{Conclusions}\label{Conclusions}

The cage model is extended to simulate DC electrophoresis, and
the drift velocity of polymers in a gel is measured as a
function of polymer length $L$ and electric field strength $E$.

The polymers behave differently in three regimes of the electric
field: in a small electric field the velocity depends linearly
on the electric field, in a high electric field the polymers are
likely to be trapped in an U-shape. The regime in between shows
a superlinear dependency on the electric field. We showed that
this dependency becomes quadratic in the electric field for
$L\to \infty$.

We have shown that the average number of kinks is not equally
distributed over the polymer in high electric fields, because
the polymers tend to get stretched in the direction of the
electric field. Hence the stored length disappears from the
polymer.

\section*{Acknowledgement}

We thank G.T. Barkema and A.G.M. van Hees for useful discussion.
The High-performance computing group of Utrecht University is
gratefully acknowledged for ample computer time.

\begin{table}
\caption{Encoding of a bond in three bits, where $x^{(i)}$ is
the $i^{\mbox{\scriptsize th}}$ bit of $x$ and so on. Note that
the encoding of the negative bonds is the binary complement of
the positive bonds.}
\label{bondencoding}
\begin{tabular}{c|ccc}
bond&$x^{(i)}$&$y^{(i)}$&$z^{(i)}$\\
\hline
\verb|+x|&$1$&$0$&$0$\\
\verb|+y|&$0$&$1$&$0$\\
\verb|+z|&$0$&$0$&$1$\\
\verb|-x|&$0$&$1$&$1$\\
\verb|-y|&$1$&$0$&$1$\\
\verb|-z|&$1$&$1$&$0$\\
\end{tabular}
\end{table}

\begin{table}
\caption{Values for $a$ and $b$, obtained by fitting the drift velocity to
the form $v^2= a E^2+ b E^4$; these values are used for scaling in figure
\ref{resultCollapse}.  For large $L$ the parameter $a$ decreases
quadratically with $L$ and $b$ is more or less constant. The graphs of
$L^2 a$ and $b$ show evidence of convergence to a constant; this is in
agreement with Equation (\ref{barkema}).}
\label{ab}
\begin{tabular}{r|r|r}
$L$&\multicolumn{1}{c|}{$a$}&\multicolumn{1}{c}{$b$}\\
\hline
30&$1.85(6)\cdot10^{-4}$&$1.9(2)\cdot10^{-2}$\\
50&$4.4(2)\cdot10^{-5}$&$5.5(2)\cdot10^{-2}$\\
70&$1.79(5)\cdot10^{-5}$&$8.3(2)\cdot10^{-2}$\\
100&$7.32(7)\cdot10^{-6}$&$1.12(1)\cdot10^{-1}$\\
140&$3.31(3)\cdot10^{-6}$&$1.29(1)\cdot10^{-1}$\\
200&$1.5(1)\cdot10^{-6}$&$1.4(7)\cdot10^{-1}$\\
\end{tabular}
\end{table}


\begin{thebibliography}{99}
\bibitem[*]{correspondent}
To whom correspondence should be addressed.
\end{thebibliography}
\end{document}